\documentstyle[12pt]{article}
\begin{document}
\centerline{\bf The basis of the Ponzano-Regge-Turaev-Viro-Ooguri
}
\centerline{\bf quantum-gravity model is the Loop Representation
basis}
\vskip 1cm
\centerline{Carlo Rovelli}
\vskip.5cm
\centerline{Physics Dept. University of Pittsburgh,
Pittsburgh PA 15260, USA}
\centerline{and Dipart. di Fisica Universita' di Trento and INFN
sez. Padova,  Italia}
\centerline{rovelli@pittvms.bitnet}
\vskip0.5cm
\centerline{\today}
\vskip 1cm
\noindent{\bf Abstract}  \vskip .5cm

We show that the Hilbert space basis that defines the Ponzano-
Regge-Turaev-Viro-
Ooguri combinatorial definition of 3-d Quantum Gravity is the same
as the one that
defines the Loop Representation. We show how to compute lengths in
Witten's 3-d
gravity and how to reconstruct the 2-d geometry from a state of
Witten's theory. We
show that the non-degenerate geometries are contained in the
Witten's Hilbert space.
We sketch an extension of the combinatorial construction to the
physical 4-d case, by
defining a modification of Regge calculus in which areas, rather than
lengths, are
taken as independent variables.  We provide an expression for the
scalar product in
the Loop representation in 4-d.  We discuss the general form of a
nonperturbative
quantum theory of gravity, and argue that it should be given by a
generalization of
Atiyah's topological quantum field theories axioms.

\vfill\eject

The problem of describing physics at the Planck scale and the
quantum
properties of gravity, is the problem of understanding what is a non-
trivial
generally covariant quantum field theory.  The last years have seen
many
developments in our understanding of these theories:  Witten's
introduction
of topological field theories, in their two versions, a' la  Chern-
Simon [1]
and a' la  Donaldson [2]; Atiyah's axiomatization of these [3];
dynamical
triangulations techniques [4]; Ashtekar reformulation of general
relativity
[5], which opened the way to the Loop Representation [6], which lead
to
discover solutions of the Wheeler-DeWitt equation, the relevance of
Knot
Theory in quantum gravity and a discrete structure of space at the
Planck
scale [7]; Turaev and Viro's [8] reformulation of the Ponzano-Regge
model [9]
in terms of quantum groups, which provides a combinatorial
definition of 3-
d topological field theories;  Crane and Vetter's [10] extension of
this
construction to 4-d; the pioneering work of Ooguri, which in 3-d has
tied the
Euclidean, combinatorial, canonical and topological definitions of
quantum
gravity [11], and in 4-d has opened the path to the Crane-Vetter
work.  These
results share a remarkable common flavor, besides, of course, the
common
long term aim of quantizing gravity. In this paper, we find the bridge
between the 3-d Ponzano-Regge-Turaev-Viro-Ooguri (PRTVO) model
and the
Loop Representation, we discuss the physical interpretation of 3-d
quantum gravity, and we sketch a general theory of physical 4-d
gravity
in which all these lines may converge.

Ponzano and Regge [9] considered Regge-calculus [13] in 3-d, but
made the
ansatz that the lengths $l_i$ of the Regge-calculus links be
constrained to be
half integers:  $l_i = j_i = 1/2 n_i$ (integer $n_i$). Half integers
$j_i$
can be interpreted
as labels of SU(2) representations. The Regge calculus action can
then be
written as a very simple expression, which is essentially a sum over
the
tetrahedra of the triangulation of the 6-j symbols of the 6 (half-
integer)
lengths $l_i=j_i$ of the links of each tetrahedron. The partition
function of
Quantum Gravity can then be constructed by fixing a sufficiently
thin
triangulation $\Delta$, and summing over its colorings c
(assignments of half
integers to every link). The reason for taking half-integer lengths,
as well
as the relation between lengths of links and SU(2) representations
appeared
to be quite mysterious at the time.  In this paper we throw some
light on
this relation.  Turaev and Viro [8] were able to show that the
Ponzano--Regge partition function is independent from the
triangulation chosen (and
transformed it in a finite sum by replacing SU(2), with a
corresponding
quantum group with a finite number of representations). Ooguri [11]
has related
the quantization of 3-d quantum gravity based on this model to the
Witten
quantization of the same theory. Ooguri construction can be
summarized in
short as follows. The quantum states of the Ponzano-Regge theory
have to
be taken, following Atiyah's general formulations of topological
quantum
field theories, as quantum combinations $\Phi_\Delta(c)$
of the colored triangulations
$(\Delta,c)$ induced on the 2-d boundary $\partial M$
of the 3-d manifold $M$. In Witten theory,
quantum states are wave functions $\Psi(\omega)$
over the moduli space of the flat
SU(2) connections $A_a^I(x)$
on the 2-d boundary ($\omega$ being equivalent classes of
 $A_a^I(x)$'s).  Ooguri relates the two representations of the theory
by
\begin{equation}
\Psi(\omega) = \sum_c \ \Phi_\Delta(c)\ \Psi_{\Delta,c}(\omega),
\end{equation}
where we have absorbed in  $\Phi_\Delta(c)$
a normalization factor appearing in
eq.(16) of Ref. [11].
The "matrix elements of the change of basis"
$$
\langle\omega|\Delta,c\rangle=\Psi_{\Delta,c}(\omega)
$$
will described in a moment. Louis Crane and Lee Smolin have
suggested
that there may be a direct relation between the Ooguri construction
and
the Loop Representation [14];
in this letter, we show, indeed, that (1) is nothing but the Loop
Transform
[6], which relates the connection representation
of Quantum Gravity with the Loop Representation. The Ooguri
representation
$\Phi_\Delta(c)$
is therefore essentially equivalent to the Loop Representation.
In the making this relation explicit, we will
provide a physical interpretation of the Ponzano-Regge ansatz
according to which the
Regge-calculus links have half-integer length and are related to
SU(2)
representations, and we will find the physical justification of the
Ooguri
construction.

The functions $\Psi_{\Delta,c}(\omega)$ introduced by Ooguri,
are constructed as follows. Given the triangulation
$\Delta$, we construct the trivalent graph dual to $\Delta$. The arc
$C_i$ of this
graph crosses the i-th link of the triangulation $\Delta$; we
associate to  $C_i$
the SU(2) representation $j_i$, where $j_i$ is the half-integer that
colors
the i-th link of $\Delta$. We then associate to $C_i$ the function
over flat SU(2)
connections $A^I_a$ given by the Wilson line $U_{j_i}[A,C_i]
= {\cal P} \exp\{\int_{C_i} A^I_a t^I_{j_i} dx^a\}$, where
the su(2) generators $t^I_{j_i}$ are taken in the $j_i$
representation.
Next, we consider the product of all these Wilson lines, where 6-j
symbols are
used to contract the indices at the trivalent intersections.
The resulting  object is a function of the triangulation $\Delta$, the
coloring $c$ and
the (flat) connection $A$.  It is gauge invariant, and thus it
defines a function over the moduli space of
the flat SU(2) connections for every  $(\Delta,c)$;
this function is $\Psi_{\Delta,c}(\omega)$ .

In order to relate this construction with the Loop Representation,
the first
observation is that, since any representation of SU(2) is obtained by
tensor
multiplication of the $j=1/2$ representation with itself, a Wilson
line
$U_{j}[A,C_i]$
in the j representation can be expressed by means of
$2j$ Wilson lines $U_{1/2}[A,C_i]$
in the 1/2
representation.  We exploit this fact by replacing
each arc $C_i$ of the trivalent graph with
precisely $2j_i$ lines, each carrying a  $U_{1/2}[A,C_i]$ parallel
transport
matrix.
Accordingly, the sum at the trivalent intersections
obtained with the 6-j symbols, can simply be replaced by the sum
over all the possible rootings of these lines at the intersection.
Finally, $\Psi_{\Delta,c}(\omega)$ can be
reexpressed as a combination of products of traces of holonomies of
A along
the resulting closed loops, all taken in the 1/2 representation.  This
follows
from elementary properties of SU(2) representation theory. In other
words,
the colored triangulation $(\Delta,c)$ uniquely determines an
ensemble
$E_{\Delta,c} = \{\alpha_1, \alpha_2,...\}$ of multiple loops (sets of
closed loops)
$\alpha_i=(\alpha_{i1}, \alpha_{i2}, ... \alpha_{iN})$,
where $\alpha_{ij}$ are
(single) loops; each multiple loop $\alpha_i$ having the
property that precisely $2j$ single
loops cross a link of the triangulation with color $j$.  The ensemble
$E_{\Delta,c}$
is defined as the set of all the homotopically inequivalent multiple
loops with this
property.  By construction we have the main relation:
$$
\Psi_{\Delta,c}(\omega) = \sum_{\alpha_i\in E_{\Delta,c}}\
\prod_i Tr U_{1/2}[A, \alpha_{ij}]
$$
Now, given a multiple loop $\alpha_i$, the product $\prod_i
Tr U_{1/2}[A, \alpha_{ij}]$ is nothing but the
loop state $|\alpha_i\rangle$, written in the connection
representation, namely
\begin{equation}
\langle A|\alpha_i\rangle = \Psi_{\alpha_i} =
\prod_i Tr U_{1/2}[A, \alpha_{ij}]
\label{loopstates}
\end{equation}
This relation is at the roots of the Loop Representation. Using this
relation, and its gauge invariance, we have
$$
\langle \omega|\Delta,c\rangle = \Psi_{\Delta,c}(\omega) =
\sum_{\alpha_i\in E_{\Delta,c}}\
\langle A|\alpha_i\rangle =
\sum_{\alpha_i\in E_{\Delta,c}}\
\langle \omega|\alpha_i\rangle,
$$
or
\begin{equation}
|\Delta,c\rangle = \sum_{\alpha_i\in E_{\Delta,c}}\
|\alpha_i\rangle.
\label{main}
\end{equation}
Eq. (\ref{main}) provides an identification between the Loop
representation basis states $ |\alpha\rangle$ and the Ooguri states
$|\Delta,c\rangle$. The relation is many-to-one because the loop
states are not independent (they form an overcomplete basis). This
relation is our first result.

In Ooguri's work, the relation (1) is postulated, and the equivalence
of the combinatorial theory with Witten's quantization is derived a
posteriori by showing the isomorphism of the two structures. Still,
the half-integer lengths remain as mysterious as they were in the
original Ponzano-Regge
paper.  To provide an interpretation of this fact, let us {\it
calculate\ }  the
lengths of the links of a triangulation in a fixed quantum state of
the
gravitational field.  A recent calculation in (3+1)-d gravity, shows
that the
area of any surface is quantized in the Loop Representation in
multiples of
1/2 (in Planck units); the area being precisely given by the number
of
intersections of the surface with the loops of the quantum state.  It
is
natural to suspect that a similar relation may work in one dimension
less.
In fact, let us show it does.  The length $l$ of a curve $C$ in 3-d
gravity is given
by
$$
l[C] = \int_C {dt \sqrt{{d C^a\over  dt}
{d C^b\over  dt} g_{ab} } } =
\int_C  dt
\sqrt{{d C^a\over  dt}
{d C^b\over  dt} E^{Ic} E^{Ic} \epsilon_{ac} \epsilon_{bd}},
$$
where $E^{Ia}$ is the variable conjugate to the connection and
$\epsilon_{ac}$
is the antisymmetric two dimensional pseudotensor.
We refer to [6] for the
notation.  In order to promote $l[C]$ to an operator, we have to deal
with the
product of the two $E$'s. Following Ref. [7], we point split the
product $E E$, by
means of the gauge invariant two-point object
$$
 Tr \left[U[A,\gamma'{}^\epsilon_x] E^a(\gamma^\epsilon_x(0))
U[A,\gamma'{}'{}^\epsilon_x] E^b(\gamma^\epsilon_x(\pi)) \right] =
=   T^{ab}[\gamma^\epsilon_x](0,\pi).
$$
Here $\gamma^\epsilon_x$ is a loop with radius
$\epsilon$, and center in $x$,
$\gamma'{}^\epsilon_x$  and $\gamma'{}'{}^\epsilon_x$ are its two
components from the value $0$ to $\pi$ and from $\pi$ to $0$
of the loop parameter, and
in the second equality we have introduced the standard Loop
Representation
notation [6] for this operator. Note that classically we have
$$
\lim_{\epsilon\rightarrow 0}  T^{ab}[\gamma^\epsilon_x](0,\pi) =
\det g  g^{ab}(x)
$$
The operator corresponding to the observable
$T^{ab}[\gamma^\epsilon_x](0,\pi)$
is well defined; using the Loop Representation, it is given [6] by
$$
\langle\alpha| T^{ab}[\gamma^\epsilon_x](0,\pi) =
= \int_\alpha ds {d\alpha^a\over ds}\delta^2 (\alpha(s),
\gamma^\epsilon_x(0) )
\int_\alpha du {d\alpha^b\over dt} \delta^2(\alpha(u),
\gamma^\epsilon_x(\pi))
\sum_i \langle\alpha\#_{su,i}\gamma^\epsilon_x |
$$
Following ref. [7], we may regularize the delta functions by a further
replacement of $\gamma^\epsilon_x$
by means of a one parameter family of loops. We can then
compute the action of the operator $l[C]$ on a loop state
$\langle\alpha|$.
The square root of
the square of the (regularized) delta function gives an absolute
value; in the
limit the intersection rearrangement gives just a multiplicative
factor, and
taking the limit we obtain, in Planck units,
$$
\langle\alpha| l[C]  =  {1\over 2}
\int_\alpha ds \int_C dt \left| {d\alpha^a\over ds} {dC^b\over dt}
\epsilon_{ab}\right| \langle\alpha|
$$
The double integral is precisely the (positive) intersection number
between $C$ and $\alpha$.
Thus we arrive at the following results: i. the length of every curve
$C$ is quantized in units of $1/2$; ii. if the gravitational field is in
the state
$\langle\alpha|$, the length of a curve $C$ is given (reinserting
conventional
units) by
$$
l[C] = n[C,\alpha] {L_{Planck}\over 2} ,
$$
where $n[C,\alpha]$ is the number of times $\alpha$ crosses $C$.

Now we can return to the PRTVO model. Result i. above implies that
{\it summing over independent states in quantum gravity means to
sum just over
quantized half-integer lengths, precisely as in the Ponzano Regge
ansatz.\ }
Moreover, the relation between the 2-dimensional colored
triangulation on
the boundaries of the manifold M and the Witten theory is now
physically transparent:
Recall that the i-th link $C_i$ of the triangulation, with has color
$j_i$ is crossed
precisely by $2j_i$ loops, thus $n[C_i,\alpha]=2j_i$; therefore,
using result ii., the
physical length in Planck units in the quantum state defined by these
loops is
$$
			l_i = 1/2 n[C,\alpha] = {1\over 2}  2j_i = j_i
$$
Thus, the physical length of the i-th link in Ooguri state is precisely
equal
to its coloring.  Therefore, {\it the state  $\Psi_{\Delta,c}(w)$,
which Ooguri associates to
the colored triangulation  $(\Delta,c)$ is a quantum state in which
the physical
length of the links of the triangulation is precisely equal to their
coloring,
in Planck units.\ }  This is our main result.

An important consequence of this result is the fact that it throws
light on the confused issue of the existence of an "unbroken phase"
and a "broken phase" in quantum gravity.  This issue concerns the
existence of states corresponding to
non-degenerate metrics in a quantization in which fields
are excitation around the zero metric configuration (as opposed to
flat).  The considerations above show that as far as 2+1 gravity is
concerned,
the problem does not sussist.  Indeed, in an generic state
$|\alpha\rangle$,
the diffeomorphism-invariant functions of the metric are
computable, and,
in general, they take values corresponding to non-degenerate metric
configurations.  The core of the subtelty is the diffeomorphism
invariance of
the theory.  The number of independent invariant observables in the
theory is finite in the quantum theory {\it as well as in the classical
theory\ }. To clarify this point, let us consider a 2-d space manifold
with
the topology of a torus.  The "no-loop" state $|0\rangle$ gives a
vanishing
metric tensor.  However, let us consider the state
$|\alpha,\beta\rangle$, where $\alpha$
and $\beta$ are the two independent non contactible loops of the
torus
(the corresponding state in the Witten representation is easily
obtained from
Eq. (\ref{loopstates})).  In this state, the minimal length of any
curve wrapping
once around the torus is ${1\over 2}L_{Planck}$.
Now let us consider the 2-d space manifolds of the classical theory.
Since spacetime if flat, we can always go to a flat euclidean 2-d
manifold by means of a gauge transformation (a 3-d
diffeomorphism). The physically meanigful information about space
is contained in two numbers that characterise a flat torus. Thus, in
this theory an arbitrary {\it non-degenerate\ }
flat space metric is characterized by these two variables. Say, for
definitness, two radii. Now let us return to the quantum theory.  In
the state $|\alpha,\beta\rangle$,
the two variables have a well-defined non-vanishing value. Thus the
quantum state $|\alpha,\beta\rangle$ corresponds to the classical
non-degenerate configuration in which space is formed by a flat
torus in which the two radii are long ${1\over 2}L_{Planck}$ each.
Bigger spaces are obtained by wrapping the loops that define the
state more times around the torus. The extrinsic curvature of these
spacelike surfaces, then, will of course be  completely
indetermined, due to Heisenberg principle. This example shows that,
at least as far as the 2+1 theory is concerned, the non-degenerate
geometries live in the same Hilbert space as the completely
degenerate state.

Before getting to the second part of this work, where we discuss the
3+1 theory, let us note that the relation
between the Loop Representation theory and the PRTVO model allows
us to  write the scalar product of two loop states $|\gamma\rangle$
and $|\gamma'\rangle$  by means of a sum
over colorings: we put $\gamma$ and $\gamma'$
on the boundaries of a (topologically trivial)
three manifold M, we fix a triangulation $\Delta$, and we have
$$
\langle\gamma|\gamma'\rangle = \sum_{c(\gamma,\gamma')}
\prod PR(c),
$$
where the sum is over all the triangulations $c(\gamma,\gamma')$
such that the coloring of
the links of the boundaries is determined by the number of times the
loops
$\gamma$ and $\gamma'$
cross the link, and by $\prod PR(c)$ we indicate the Ponzano-Regge
product of the
coloring.

The above result indicates a direction for constructing the physical
4-d
theory.   Let us consider the 4-d manifold M, with, say, two
boundaries
$\partial M_1$  and $\partial M_2$.
We begin by fixing a 4-d triangulation $\Delta$ of spacetime, which
induces 3-d triangulations of the two boundaries. In 4-d, they are
the areas
of surfaces, not the lengths, that are naturally quantized in 1/2 the
Planck
unit.  Thus, it is natural to use as independent variables for Regge
calculus
the areas of the faces, rather than the lengths of the links.  A key
observation is that a 4-d simplex has the same number (10) of faces
(2-d
symplices) and links (1-d simplicies).   Therefore we can generically
invert
the relation between lengths and areas, and express the lengths of
the 10
links of each 4-simplex as functions of the areas of the 10 faces.
Let
$a_1\ ... \ a_{10}$
be the areas of the 10 faces of a fixed 4-simplex $s$.  The Regge
action of
the simplex can be expressed as a function of these areas:
$S_{Regge}(s) =
S_{Regge}(a_1\ ... \ a_{10})$.
Note that $S_{Regge}$  must be a function of 10 variables with the
full symmetry of the 4-simplex. We suspect that the corresponding
quantity
$S_{Regge}(a_1\ ... \ a_{10})$, seen as function of half-integer
variables, has an
interpretation in terms of group representation theory (at least in
the large
$a_i$ limit) analogous to the 6-j symbols interpretation of its 3-d
analog, but
we have not found it.
\footnote
{The reason the group SU(2) is still the relevant group, in spite of
the
fact we are one dimension above, is in the very roots of Ashtekar's
magic construction: the (complexified) Lorentz group splits
naturally in two
(complexified) SO(3) groups, its self-dual and antiselfdual parts,
and, as shown by
Ashtekar, the full theory can be constructed using only one of the
two SO(3)
components. }
The above construction defines a combinatorial
quantum theory in 4-d (for a fixed triangulation). In the absence of
boundaries, we have
$$
Z(\Delta)= \sum_c Z_{Regge}(\Delta,c) =
\sum_{c\ {\rm of}\ \Delta} \prod_s \exp\{S_{Regge}(c)\}
$$
where the sum is over the colorings, the product over the 4-
simplices.  The
states of this theory are given by the induced colorings of the
induced
triangulation on the 3-d boundary of the 4-d triangulated spacetime.
These
states are physically determined by the fact that there is a three
dimensional triangulation of space such that the (2-d) surfaces of
the
triangulation have assigned areas; these areas being half-integers in
Planck
units. These states are therefore in correspondence with the states
of the
Loop Representation. The correspondence being given by the result of
ref.[7],
that the physical area of any surface is 1/2 the number of loops that
cross
the surface.   Since the states can be written in terms of loop
states, and
viceversa, this theory defines a scalar product in loop space by:
$$
\langle\gamma|\gamma'\rangle_\Delta =
\sum_{c(\gamma,\gamma'){\rm of}\ \Delta} Z_{Regge}(\Delta,c) ,
$$
where $ \gamma$ is on $\partial M_1$ and  $\gamma'$ is on
$\partial M_2$,
and the sum is over all the colorings of
the interior areas, the colorings of the areas on the boundaries being
determined as 1/2 the number of times the loops cross the surface.
The
idea that states of the Loop Representation can be better understood
in
terms of the area they induce on a 3-d triangulation was
proposed by Lee Smolin [15].

We still do not have a well defined diffeomorphism invariant theory,
since
the above construction depends on the triangulation $\Delta$.
\footnote{An optimistic hope would be that the scalar product above
does not
depend on the triangulation. While we do not hope for so much, still
we do not think
this be totally impossible, as some considerations below may
suggest.}
We eliminate this dependence by summing over all the triangulations
of M.   The key
point in this construction is that this is possible to sum over the
triangulations,
because we have a way of naming the
quantum states in the boundaries, which is independent from the
triangulation of the boundary itself, namely the loop basis. Thus, we
define
a theory as
\begin{equation}
Z = \sum_\Delta \sum_{c\ {\rm of}\ \Delta}  Z_{Regge}(\Delta,c)
\label{theory}
\end{equation}
and the scalar product between two loop states by
\begin{equation}
\langle\gamma|\gamma'\rangle =
\sum_\Delta  \langle\gamma|\gamma'\rangle_\Delta
\label{product}
\end{equation}
(If in Eq. (\ref{theory}) we perform the sum over the colorings first,
the theory
takes a form strictly related to the dynamical triangulations
approaches to quantum
gravity [4]; this relation, we believe, deserves to be better
explored.)  The
important point is that scalar product defined in this way is
invariant under
independent diffeomorphisms on each of the two loops, because so is
the
sum.  Therefore it defines a scalar product $\langle K|K'\rangle$
between knot states by
$$
\langle K(\gamma)|K(\gamma')\rangle =
\langle\gamma|\gamma'\rangle
$$
These equations provide a formal expression for the Hilbert
structure of
quantum gravity. We expect that Eq. (\ref{product}) defines a
projector on the knot
states, which projects on the solutions of the Wheeler DeWitt
equation, as
it happens in 3-d [11].  This is our main proposal for a 4-d theory.

To clarify the meaning of Eqs. (\ref{theory}) and (\ref{product}),
let us rewrite them in terms
of the original continuous Ashtekar's variables $A$ and $E$. We have
\footnote{
Ooguri suggests [12] that in a one may have two conjugate variables
in the continuum
version of the theory, which correspond respectively to the choice of
the
triangulation and the choice of the coloring in the combinatorial
version of the theory.
Note the similarity between the  Ashtekar's action $S[A,E]=\int
F\wedge E \wedge E$,
with the (topological !) BF theory action $S[A,B]=\int F\wedge B$
which seems to underlie the Ooguri-Crane-Vetter invariants [10,12].
The relation
between the construction proposed here and the (triangulation
independent) Ooguri-
Crane-Vetter construction deserves to be studied in detail .
}
$$
\langle\gamma|\gamma'\rangle  =
\int [d^4A] [d^4E]  \exp\{-1/\hbar S_E[^4A,^4E]\} {\Psi_\gamma(A)}
\Psi^*_{\gamma'}(A)
$$
where
$$
\Psi_\gamma(A)=   \Psi_\gamma(^3A) = \langle^3A|\gamma\rangle =
tr {\cal P} \exp\oint_\gamma A
$$
and each loop is in one of the two boundaries of spacetime.  It is
easy to
convince oneself that this is the correct formal expression for the
scalar
product in the Loop Representation, up the problem of definition and
finiteness of the functional integration. To our knowledge, this
expression
was first suggested by Maurizio Martellini [16].  This is the
connection
representation analog of Hawking expression [17]
$$
\langle\Psi|\Phi\rangle   = \int [d^4g] \exp\{-1/\hbar S_E[^4g]\}
{\Psi^*(^3g)} \Phi(^3g')
$$
where the integration is over all the 4-d metrics ${}^4g$ on M, and
${}^3g$  and ${}^3g'$ are
the restrictions of ${}^4g$ to the two components of the boundary of
M.  Of course
these functional integrals do not mean anything until a definition
is provided; in particular, if we want to compute them in a
perturbation
expansion around flat space, we encounter the weel known
gravitational divergences.
The combinatorial expression given above is a proposal for
this definition.
\footnote{
Of course, each of the above equations can be immediately
generalized to non-trivial
topologies, Hartle-Hawking states, disconnected universes, and so
on, if one is
interested in this kind of physics directions.
}
In this continuum case, Hawkings does indeed give a formal
derivation of the fact that the functional integral defines a
projection on
the solutions of the Wheeler DeWitt equation. This is an indirect
support of
our expectation that Eq.(\ref{product}) projects on the solutions of
the
hamiltonian constraint.

We conclude with a consideration on the formal structure of 4-d
quantum
gravity, which is important to understand the above construction.
Standard
quantum field theories, as QED and QCD, as well as their
generalizations like
quantum field theories on curved spaces and perturbative string
theory, are
defined on metric spaces.  Witten's introduction of the topological
quantum
field theories has shown that one can construct quantum field
theories
defined on a manifold which has only its differential structure, and
no fixed
metric structure. The theories introduced by Witten and axiomatized
by
Atiyah have the following peculiar feature: they have a finite number
of
degrees of freedom, or, equivalently, their quantum mechanical
Hilbert
spaces are finite dimensional; classically this follows from the fact
that
the number of fields is equal to the number of gauge
transformations.
However,
not any diff-invariant field theory on a manifold has a finite number
of
degrees of freedom.  Witten's gravity in 3-d is given by the action
$S[A,E] = \int F \wedge E$,
which has finite number of degrees of freedom. Consider the action
$S[A,E] =  \int F \wedge E \wedge E$, in 3+1 dimensions, for a (self
dual) SO(3,1)
connection $A$ and a (real) one form $E$ with values in the vector
representation of
SO(3,1). This theory has a strong resemblance with its 2+1
dimensional analog: it is
still defined on a differential manifold without any fixed metric
structure,
and is diffeomorphism invariant.  We expect that a consistent
quantization
of such a theory should be found along lines which are more
similar to the quantization of the $\int F \wedge E$,
 theory than to the quantization of
theories on flat space, based on the Wigthman axioms namely on n-
points
functions and related objects.  Still, the theory $\int F \wedge E
\wedge E$
has genuine field
degrees of freedom: its physical phase space is infinite dimensional,
and we
expect that its Hilbert state space will also be infinite dimensional.
There
is a popular belief that a theory defined on a differential manifold
without metric and diffeomorphism invariant has necessarily a
finite
number of degrees of freedom ("because thanks to general covariance
we can
gauge away any local excitation"). This belief is of course wrong.  A
theory
as the one defined by the action $\int F \wedge E \wedge E$
is a theory that shares many
features with the topological theories, in particular, no quantity
defined "in
a specific point" is gauge invariant; but at the same time
it has genuinely infinite degrees of
freedom.  Indeed, this theory is of course nothing but (Ashtekar's
form of)
standard general relativity.

The fact that "local" quantities like the n-point functions are not
appropriate to describe quantum gravity non-perturbatively has been
repeatedly noted in the literature.  As a consequence, the issue of
what are
the quantities in terms of which a quantum theory of gravity can be
constructed is a much debated issue. The above discussion indicates
a way to
face the problem: The topological quantum field theories studied by
Witten
and Atiyah provide a framework in terms of which quantum gravity
itself
may be framed, in spite of the infinite degrees of freedom.  In
particular,
Atiyah's axiomatization of the topological field theories provides us
with a
clean way of formulating the problem.  Of course, we have to relax
the
requirement that the theory has a finite number of degrees of
freedom.
These considerations leads us to propose that the correct  general
axiomatic
scheme for a physical quantum theory of gravity is simply Atiyah's
set of
axioms [3] up to finite dimensionality of the Hilbert state space.
We denote
a structure that satisfies all Atiyah's axioms, except the finite
dimensionality of the
state space,  as a {\it generalized topological theory\ }.

The theory we have sketched is an example of such a generalized
topological
theory. We associate to the connected
components $\partial M_i$ of the boundary of M the infinite
dimensional state space of the Loop Representation of quantum
gravity.
Eq.(\ref{product}), then, provides a map,
in Atiyah's sense, between the state spaces constructe on two of
these boundary
components. Equivalently, it provides the definition of the Hilbert
product in
the state space.

One could argue that the framework we have described cannot be
consistent, because it cannot allow us to recover the "broken phase
of gravity" in which we have a nondegenerate background metric: in
the proposed framework one has only non-local observables on the
boundaries, while in the broken phase
a {\it local\ } field in M has non-vanishing vacuum expectation value.
We think that this argument is weak because it disregards
the diffeomorphism invariance of the theory: in classical general
relativity no experiment can distinguish a Minkowskian spacetime
metric from a non-Minkowkian flat metric.  The two are physically
equivalent, as two gauge-related Maxwell potentials.  For the same
reason,
no experiment could detect the absolute {\it position\ } of, say, a
gravitational wave,
(while of course the position of an e.m. wave is observable
in Maxwell theory).  Physical locality in general relativity is only
defined as coincidence of some physical variable with some other
physical variable, while in non general relativistic physics locality
is
defined with respect to a fixed metric structure.
In classical general relativity, there is no
gauge-invariant obervable which is local in the coordinates. Thus,
any observation can be described by means of the value of the fields
on arbitrary boundaries of spacetime. This is the correct
consequence
of the fact that "thanks to general covariance we can
gauge away any local excitation", and this is the reason for which
one
can have the ADM "frozen time" formalism.
The spacetime manifold of general relativity is, in a sense, a much
weaker
physical object than the spacetime metric manifold of ordinary
theories.
All the general relativistic physics can be read from the boundaries
of this
manifold. At the same time, however, these boundaries still carry an
infinite
dimensional number of degrees of freedom.

Finally, we must recall that the computation of the evolution of
expectation
values in physical time (as opposed to coordinate time, which has no
diffeomorphism invariant meaning) requires the use a physical clock
coupled
to the theory (in principle this could also be a component of the
gravitational field itself) [18]. In this sense the integration (or the
sum)
over the M is physically very analogous to the derivation of the
propagator
of a relativistic particle by means of an integral over the paths
$x^\mu(\tau)$,
where $\mu=0,1,2,3$;
in the particle case too, indeed, the scalar product between two
wave
functions on Minkowski space can be obtained by integrating over a
five
dimensional manifold that interpolates two Minkowski spaces (see
for
instance [19]). Physical evolution, of course, is in $x_o$, not in
$\tau$,
namely in 4-d, not in 5-d.  In addition, we should also recall that
the quantization of the physical area is a non-gauge invariant result,
unless
reinterpreted in some suitably gauge-invariant context [20].

Summarizing, we have have shown the following:

i. The "colored
triangulation" basis of the Ponzano-Regge-Turaev-Viro-Ooguri
quantization
of 3-d gravity is precisely the Loop Representation basis.

ii. In Witten's theory, we can compute the lengths of the arcs of a
Regge
traingulation, (up to diffeomorphisms).

iii. We can
interpret the quantization of the length in half integer units in
physical
terms: the spectrum of the length operator has discrete half-integer
eigenvalues.

iv. These lengths are related to the SU(2) representations
because the quantum states that diagonalize the lengths are given
(in the
connection representation) by Wilson lines that cross the curve 2j
times, or,
equivalently, by one Wilson line in the j representation that crosses
the
curve.

Motivated by these results, we have sketched a 4-d combinatorial
theory, based on a modification of Regge calculus.   Many questions
remain
open as far as this theory is concerned, the most relevant ones being
the
relation between the Euclidean and the Lorentzian theory, and the
convergence properties of the sum (3).   Finally, we have proposed
that
Atiyah's axiomatization of topological field theories can be extended
also to
theories with infinite degrees of freedom, and that this extension
can be
takes as the general formal structure of a quantum theory of gravity.

\vskip 1cm
\vskip 1cm
\vskip 1cm
\bf References
\rm
\vskip 1cm

[1] E Witten, Comm Math Phys 121 (1989) 351; Nucl Phys B322
(1989) 622; B330
(1990)
285;  B311 (1988/89) 46; Nucl Phys B340 (1990) 281

[2] E Witten, Comm Math Phys 117 (1988) 353

[3] MF Atiyah, The Geometry and Physics of Knots , Accademia
Nazionale dei Lincei,
Cambridge
University Press 1990; Publ Math Inst hautes Etudes Sci Paris 68
(1989) 175

[4] J Ambjorn, B Durhuus, T Jonsson, Mod Phys Lett A6 (1991) 1133;
J Ambjorn, Nucl
Phys
B25A (1992) 8; J Ambjorn, J Jurkiewicz Phys Lett B278 (1992) 42.
ME Agishtein, AA
Migdal,
Mod Phys Lett A6 (1991) 1863. Godfrey, M Gross Phys Rev D43
(1991) R1749; M Gross,
Nucl
Phys B20 (1991) 724

[5] A Ashtekar, Phys Rev Lett 57 (1986) 2244; Non-perturbative
canonical gravity,
Lecture
notes in collaboration with RS Tate, World Scientific 1991

[6] C. Rovelli, L. Smolin, Phys Rev Lett 61 (1988) 1155; Nucl Phys
B331 (1990) 80. For
a
review, see: C Rovelli, Class Quant Grav 8 (1991) 1613. For the 2+1
dimensional
theory see: A
Ashtekar, V Husain, C Rovelli, J Samuel, L Smolin, Class and Quant
Grav 6 (1989) L185

[7]  A. Ashtekar, C. Rovelli, L. Smolin, Phys Rev Lett 69 (1992) 237

[8]  V Turaev, O Viro, Topology 31 (1992) 865

[9] G Ponzano, T Regge, "Semiclassical limits of Racah Coefficients"
in Spectroscopy
and Group
theoretical methods in Physics. Ed F Bloch, North Holland,
Amsterdam 1968

[10] L Crane, D Yetter, "A categorical construction of 4D topological
quantum field
theories",
Kansas State University preprint 1992

[11] H Ooguri, Nucl Phys B382 (1992) 276

[12] H Ooguri, Mod Phys Lett A7 (1992) 2799

[13] T Regge, Nuovo Cimento 19 (1961) 551

[14] L Smolin, personal communication; L Crane, "Categorical
Physics", Kansas State
prep 1992

[15] L Smolin, "Time, measurement and information loss in quantum
cosmology",
Syracuse
preprint 1993

[16] M Martellini, personal communication

[17] SW Hawking, in General Relativity, an Einstein centenary Survey
, eds. SW
Hawking, W
Israel, Cambridge University Press 1979

[18] C Rovelli, Class and Quant Grav 8 (1991) 297; 8 (1991) 317;
Phys Rev D42 (1991)
2638; D43 442 (1991)

[19] PD Mannheim, Phys Lett 166B (1986) 191

[20] C. Rovelli, "A Generally covariant quantum theory and a
prediction on quantum
measurement
of geometry", sumbitted to Nucl Phys (1992).

\end{document}